\documentclass[mathleft,a4paper
]{an}
\usepackage{graphicx}
\usepackage[english]{babel}
\usepackage{times}
\usepackage{amsmath}
\usepackage[]{natbib}
\usepackage{array}
\usepackage{txfonts}
\bibpunct{(}{)}{;}{a}{}{,}
\def\aj{AJ}%
\def\apj{ApJ}%
\def\apjl{ApJ}%
\def\apss{Ap\&SS}%
\def\aap{A\&A}%
\def\aaps{A\&AS}%
\def\mnras{MNRAS}%
\begin{document}

\Pagespan{789}{}
\Yearpublication{2006}%
\Yearsubmission{2005}%
\Month{11}%
\Volume{999}%
\Issue{88}%

\title{J1128+592: a highly variable IDV source}

\author{K.~\'E. Gab\'anyi\inst{1,2}\fnmsep\thanks{Corresponding author:
  \email{gabanyik@sgo.fomi.hu}\newline}
\and N. Marchili\inst{3}
\and T.~P. Krichbaum\inst{3}
\and S. Britzen\inst{3}
\and L. Fuhrmann\inst{3}
\and A. Witzel\inst{3}
\and J.~A. Zensus\inst{3}
\and P.~M\"uller\inst{3}
\and X. Liu\inst{4}
\and H.~G. Song\inst{4}
\and J.~L. Han\inst{5}
\and X.~H. Sun\inst{5}
}
\titlerunning{J1128+592: a highly variable IDV source}
\authorrunning{K.~\'E. Gab\'anyi et al.}
\institute{
Hungarian Academy of Sciences, Research Group for Physical Geodesy and
Geodynamics, Budapest, Hungary
\and
F\"OMI Satellite Geodetic Observatory, Budapest, Hungary
\and 
Max-Planck-Institute f\"ur Radioastronomie, Bonn, Germany
\and
Urumqi Observatory, National Astronomical Observatories, Chinese
Academy of Sciences, Urumqi 830011, PR China 
\and National Astronomical Observatories, Chinese Academy of Sciences, Beijing 100012, PR China}

\received{}
\accepted{}
\publonline{later}


\abstract{%
Short time-scale radio variations of compact extragalactic radio quasars and
blazars known as IntraDay Variability (IDV) can be explained in at least some
sources as a propagation effect; the variations are interpreted as
scintillation of radio waves in the turbulent interstellar medium of the Milky
Way. One of the most convincing observational arguments in favor of a
propagation-induced variability scenario is the observed annual modulation in
the characteristic time scale of the variation due to the Earth's orbital
motion. So far there are only two sources known with a well-constrained seasonal cycle.
Annual modulation has been proposed for a few other less well-documented
objects. However, for some other IDV sources source-intrinsic structural 
variations which cause drastic changes in the variability time scale were also 
suggested. J1128+592 is a recently discovered, highly variable IDV source. Previous,
densely time-sampled flux-density measurements with the Effelsberg 
100-m radio telescope (Germany) and the Urumqi 25-m radio telescope 
(China), strongly indicate an annual modulation of the
time scale. The most recent 4 observations in 2006/7, however, do not fit well
to the annual modulation model proposed before. In this paper, we investigate 
a possible explanation of this discrepancy.}

\maketitle

\section{Introduction}

Short time-scale (few hours to few days long) variations of the radio flux 
density in flat
spectrum radio-loud quasars and blazars were discovered in the mid-eighties \citep{idv_discovery,idv_discovery2}. The phenomenon was named
IntraDay Variability (IDV). If interpreted as being source intrinsic, 
the short time scale of the 
variations would imply -- through the light travel time argument -- micro-arcsecond-scale sizes 
of the emitting regions, which would result in excessively large apparent brightness
temperatures. The calculated variability brightness temperatures obtained are typically 
in the range of $10^{16} \text{\,K to } 10^{21}$\,K, which is far 
in excess of the inverse-Compton limit of $\sim 10^{12}$\,K \citep{compton}.
Thus, theories which explain IDV with variations intrinsic to the quasar,
require excessively large Doppler boosting factors ($\delta$\,$\gg$\,$50$) or 
special source geometries \citep[such as non-spherical relativistic emission models,\\e.g.][]{qian_0716,qian_injection,qian_0917,spada} or 
coherent and collective plasma emission \citep{coherent_benford,coherent_lesch} 
to avoid the inverse-Compton catastrophe.

An alternative theory explains IDV as a propagation effect. In this source-extrinsic 
interpretation, IDV is caused by interstellar 
scintillation (ISS) of radio waves in the turbulent plasma of the Milky
Way. 
One of the most convincing arguments in favor of an extrinsic explanation of
IDV is the so called annual modulation of the IDV time scale. The
characteristic variability time scale is inversely proportional to the relative velocity between the observer and the
scattering medium. The observer's velocity (and so the relative velocity vector
between the observer and scattering medium) undergoes a systematic 
annual modulation as the Earth orbits around the Sun. This annual velocity variation
then is observed as an annual change in the variability time scale.
Such seasonal cycles are seen in two IDV sources: J1819+3845
\citep{annual1819} and PKS1257-326 \citep{time_delay3}. 
In a few other IDV sources, such as PKS1519-273
\citep{annual1519_1},\\B0917+624 \citep{0917annual1, 0917annual2} PKS0405-385 \citep{0405annual} and B0954+658 (Fuhrmann priv. com.), 
different characteristic variability time scales were measured at 
different observing epochs, however systematic seasonal variations, as expected
from the annual modulation model, were either not seen or could not be 
unambiguously identified. The IDV sources\\B0917+624 \citep{cease_0917} and PKS0405-385 seem to display the so 
called episodic IDV \citep[e.g.\\][]{0405annual},
where pronounced variability is observed for months or even years
and then temporarily stops and the IDV ceases. 
This variability behaviour may be due to changes in the intrinsic 
source structure (expansion of the scintillating component) or 
changes in the scattering plasma
\citep{cease_0917_2,cease_0917_3,cease_0917_4,0405annual}. 

If IDV is caused by ISS, it is possible to predict the frequency dependence
of the variability amplitude, which above a transition frequency from strong to
weak scintillation is expected to be less pronounced towards higher
observing frequencies. However, observations of several IDV
sources \citep[most notably S50716+71,][]{reduc_idv,agudo_0716,ostorero_0716}
do not show such a frequency dependence and show variability amplitudes which
increase with frequency. \cite{cease_0917_3} suggested that 
a source-intrinsic contribution may cause the discrepancy between the observed 
and the expected frequency dependence.

For a better understanding of the IDV phenomenon, a clear distinction 
between ISS induced propagation effects and a source-intrinsic origin of the
variability is important. Here, we present the variability characteristics of the recently discovered 
highly variable IDV source J1128+592 and discuss the possible mixture 
of ISS-induced and source-intrinsic variability.

\section{Observations}

J1128+592 was observed with the Max-Planck-Institut f\"ur Radioastronomie (MPIfR) 100-m Effelsberg radio telescope 
(at 2.70, 4.85 and 10.45\,GHz) and with the Urumqi radio telescope (at
4.85\,GHz) in so far 15 observing sessions, each lasting several days. 
The results of the first ten epochs of observations are already published in
\cite{1128_first}.

The low-noise 4.85\,GHz receiver, a new receiver backend and new telescope driving software 
for the Urumqi telescope were provided by the MPIfR.
A detailed technical description of the receiver system and the telescope is
given in e.g. \cite{urumqi_tel}.

The observing epochs are irregularly distributed
over a $\sim 2$-year interval between December 2004 and January 2007. 
The 4.85\,GHz observing epochs, and the observing radio 
telescopes are listed in Col. 1 and  Col. 2 of Table \ref{tab:obs1128}.

\begin{table}
\caption{\label{tab:obs1128}Summary of the 4.85\,GHz IDV observations of J1128+592. The
  table lists the observing dates (Col. 1), the observing radio telescopes
  (Col. 2, 'E' for Effelsberg, 'U' for Urumqi), the average
  modulation index of the calibrators ($m_0$, Col. 3), the modulation index of
  J1128+592 (Col. 4) and the derived characteristic variability time scale 
  (Col. 5) in days.}
\centering
\begin{minipage}{7.9cm}
\renewcommand{\thefootnote}{\thempfootnote}
\begin{tabular}{cc>{$}c<{$}>{$}c<{$}>{$}c<{$}}
\hline
Epoch & R.T. & m_0 & m & t_\text{scint} \\
 & & (\%) & (\%) & \text{(day)} \\
\hline
25-31.12.2004\footnote{\cite{1128_first}} & E & 0.4 & 10.9 & 0.30 \pm 0.10 \\
13-16.05.2005\footnotemark & E & 0.5 & 2.2 & 0.88 \pm 0.10 \\
14-17.08.2005\footnotemark & U & 0.6 & 5.9 & 0.90 \pm 0.30 \\
16-19.09.2005\footnotemark & E & 0.6 & 2.9 & 0.55 \pm 0.15\\ 
27-31.12.2005\footnotemark & U & 1.2 & 7.9 & 0.35 \pm 0.08\\ 
29-30.12.2005\footnotemark & E & 0.4 & 7.2 & 0.34 \pm 0.06\\ 
10-12.02.2006\footnotemark & E & 0.4 & 6.8 & 0.10 \pm 0.05\\ 
15-18.03.2006\footnotemark & U & 0.5 & 5.7 & 0.37 \pm 0.10 \\
28.04-02.05.2006\footnotemark & E & 0.5 & 9.0 & 1.52 \pm 0.15 \\
28.04-02.05.2006\footnotemark & U & 0.5 & 7.0 & 1.51 \pm 0.25 \\
10-13.06.2006\footnotemark & U & 0.5 & 4.1 & 0.51 \pm 0.08 \\
14-18.07.2006\footnotemark & U & 0.7 & 5.8 & 0.60 \pm 0.25 \\
19-25.08.2006 & U & 0.7 & 4.7 & 1.20 \pm 0.20 \\
23-27.09.2006 & U & 0.8 & 2.3 & 1.30 \pm 0.40 \\
16-18.11.2006 & U & 0.6 & 4.7 & 1.10 \pm 0.10 \\
16-18.12.2006\footnote{There were not enough measurements to calculate a
  reliable time scale.} & E & 0.5 & 3.0 & - \\
13-15.01.2007\footnote{Due to human error, slightly different frequency setup
  was used than previously. The center frequency was 4.79\,GHz instead of
  4.85\,GHz.} & E & 0.4 & 5.2 & 0.40 \pm 0.14 \\
\hline
\end{tabular}
\end{minipage}
\end{table}

At both telescopes, the flux density measurements were performed 
using the cross-scan technique, in which the telescope is moved repeatedly 
(4 -- 8 times) in azimuth and in elevation 
over the source position. Each such movement is called a subscan. After 
baseline subtraction, a Gaussian curve was fitted for each subscan to the resulting 
slice across the source. For each scan, the averaged and pointing corrected 
peak amplitude of the Gaussian curve yielded a measure of the source flux density. Then the systematic elevation and time-dependent gain variations were corrected, using the combined gain curves and gain-transfer functions of calibrator sources 
of known constant flux density. The measured flux densities were then tied to the absolute flux-density scale.
The absolute flux-density scale was de\-ter\-mined from repeated observations of the primary calibrators e.g. 
\object{3C\,48}, \object{3C\,286}, \object{3C\,295} and \object{NGC\,7027} and
using the flux-density scale of \cite{abs1} and \cite{abs2}.

A more detailed description of the data reduction and the comparison of
accuracy reached by the two telescopes is given in \cite{1128_first}.

\subsection{Data analysis} 

For the variability analysis, we follow the method introduced by \cite{idv_discovery2} 
and described in detail in e.g. \cite{reduc_idv2}, \cite{reduc_idv}.

To decide whether a source can be regarded as variable, we performed 
a $\chi^2$ test.
We tested the hypothesis whether the light curve can be modeled 
by a constant function. The sources for which 
the probability of this hypothesis was less then 0.1\,\% were considered to 
be variable. The amount of variability in a light curve is described by the 
modulation index and the noise-bias corrected variability amplitude. The 
modulation index is defined as $m=100\cdot\frac{\sigma}{\langle \text{S} \rangle}$, where 
$\langle \text{S} \rangle$ is the time average of the measured flux density, and $\sigma$ is the standard 
deviation of the flux density. The average modulation index of the calibrator sources
($m_0$) provides a conservative estimate of the overall calibration accuracy. It
describes the amount of residual scatter in the data set after all the
calibration steps are performed. The $m_0$ and $m$ of J1128+592 are listed
in Col. 3 and in Col. 4 of Table \ref{tab:obs1128}, respectively.

To determine the characteristic variability time scale, we made
use of the structure function \citep[for a definition see e.g. ][]{sf_def}, the autocorrelation
function and the light curve itself. 
We defined the characteristic variability time scale by the time-lag, where the 
structure function reaches its saturation level. This corresponds to the first minimum of the
autocorrelation function. We compared\\these with the average peak-to-trough
time derived from the light curve \citep[for details see ][]{1128_first}. In Col. 5 of Table \ref{tab:obs1128}, we list
the variability time scales obtained by the average from the three different methods.

\section{Results and discussion}

\begin{figure*}
 \begin{minipage}[t]{0.33\textwidth}
 \begin{center}
  \includegraphics[width=5.65cm]{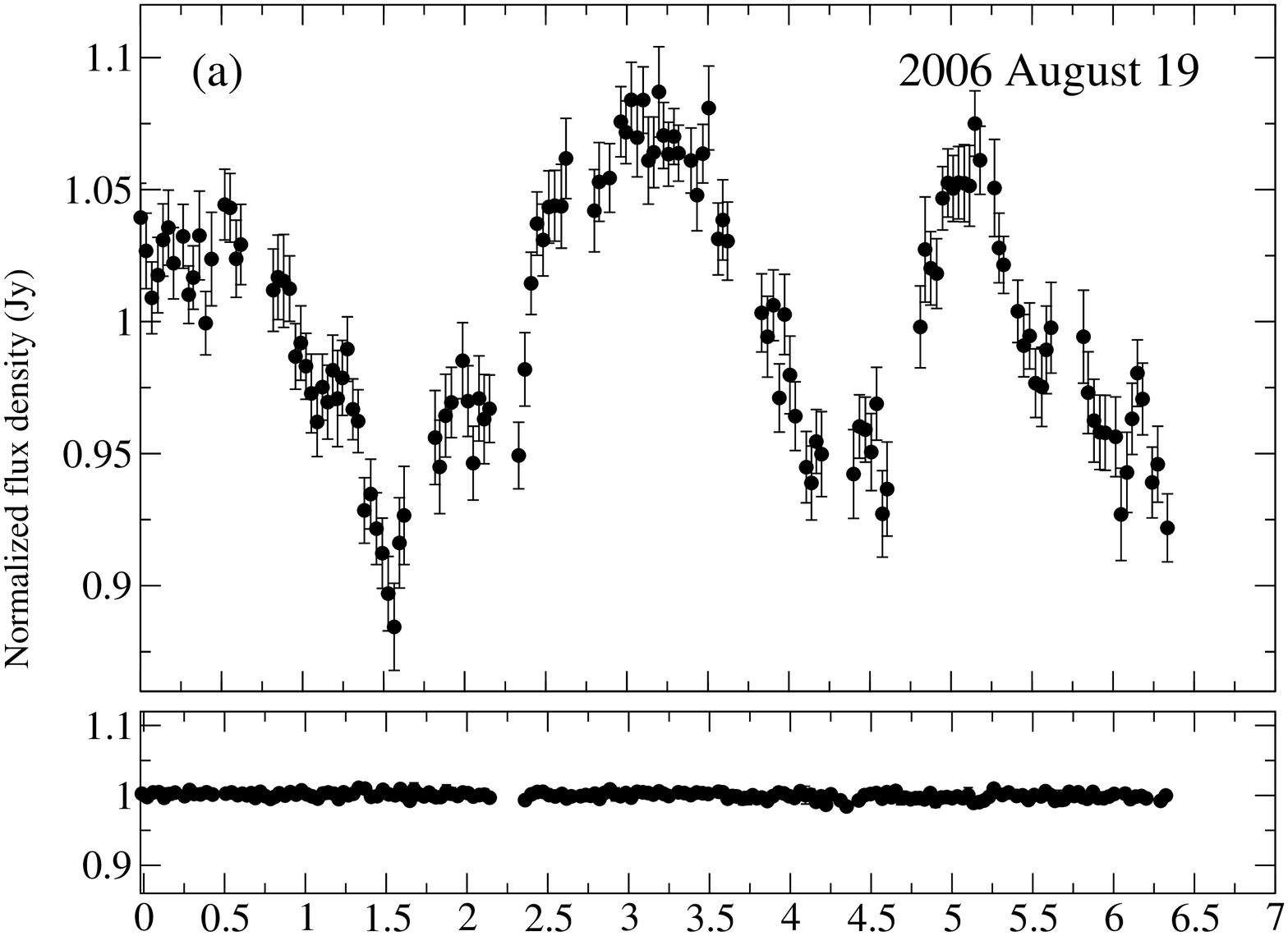}
 \end{center} 
 \end{minipage}
 \hfill
 \begin{minipage}[t]{0.33\textwidth}
 \begin{center}
  \includegraphics[width=5.65cm]{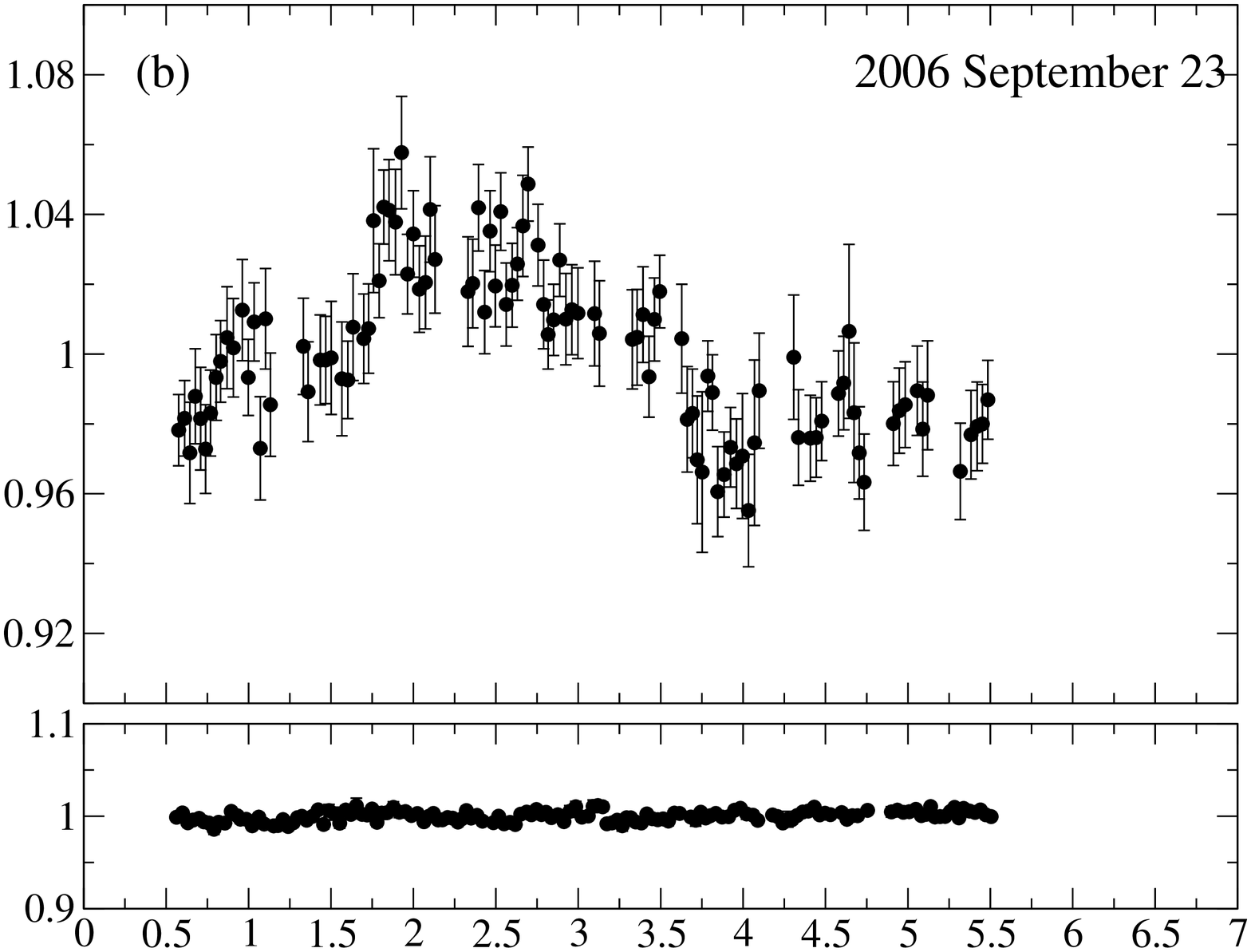}
 \end{center}
 \end{minipage}
\hfill
 \begin{minipage}[t]{0.33\textwidth}
 \begin{center}
  \includegraphics[width=5.65cm]{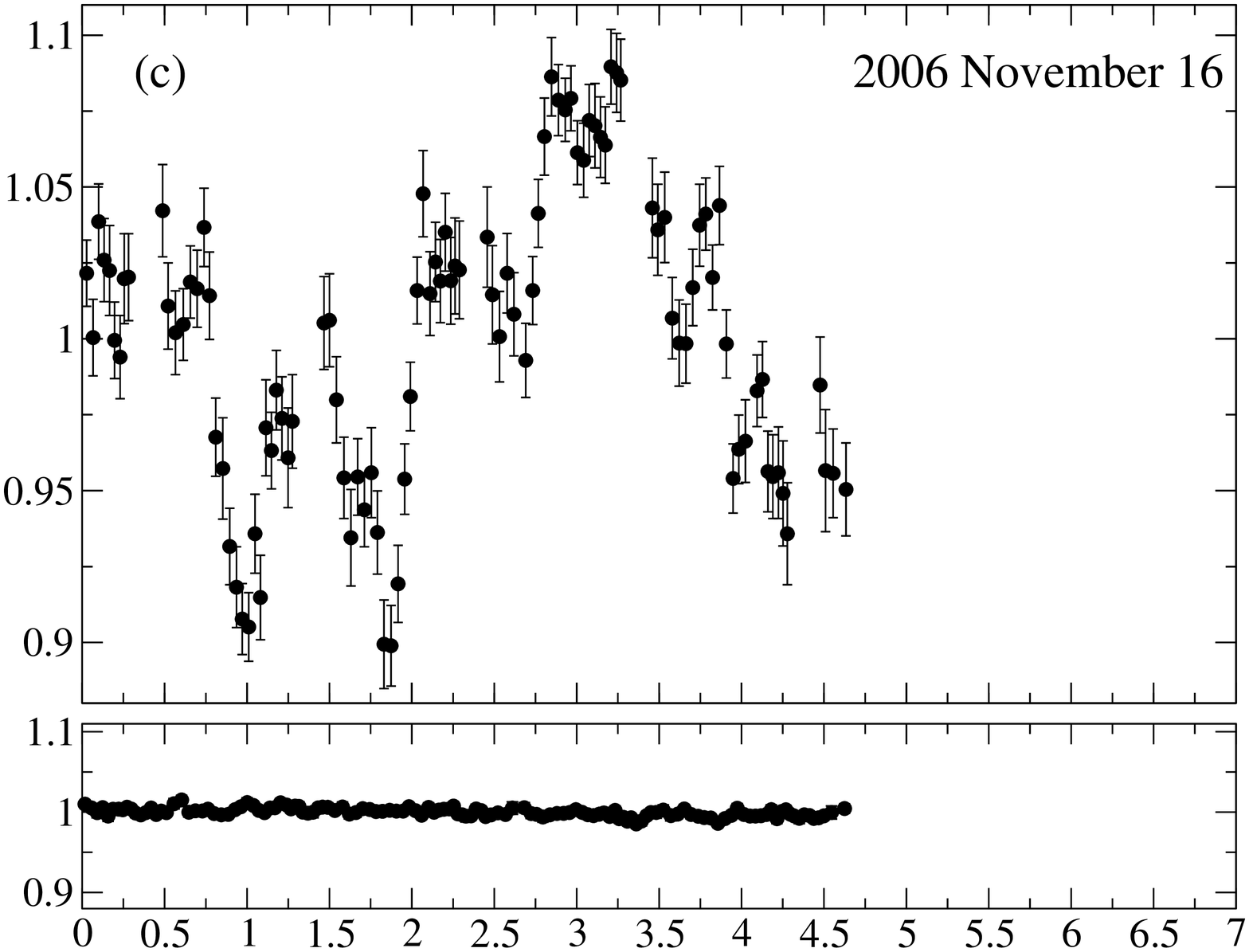}
 \end{center} 
 \end{minipage}
 \begin{minipage}{0.33\textwidth}
 \begin{center}
  \includegraphics[width=5.65cm]{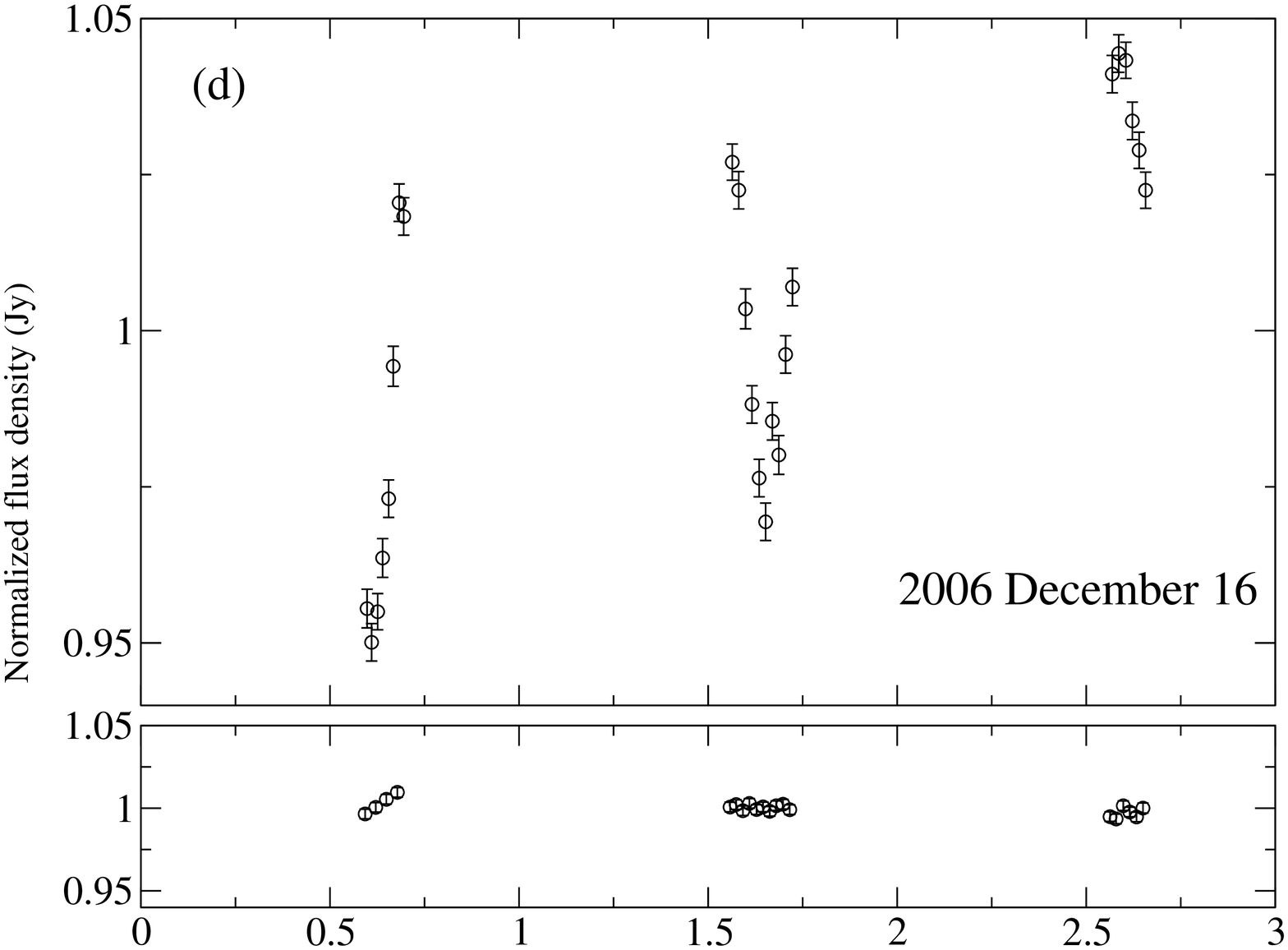}
 \end{center}
 \end{minipage}
 \begin{minipage}{0.33\textwidth}
 \begin{center}
  \includegraphics[width=5.65cm]{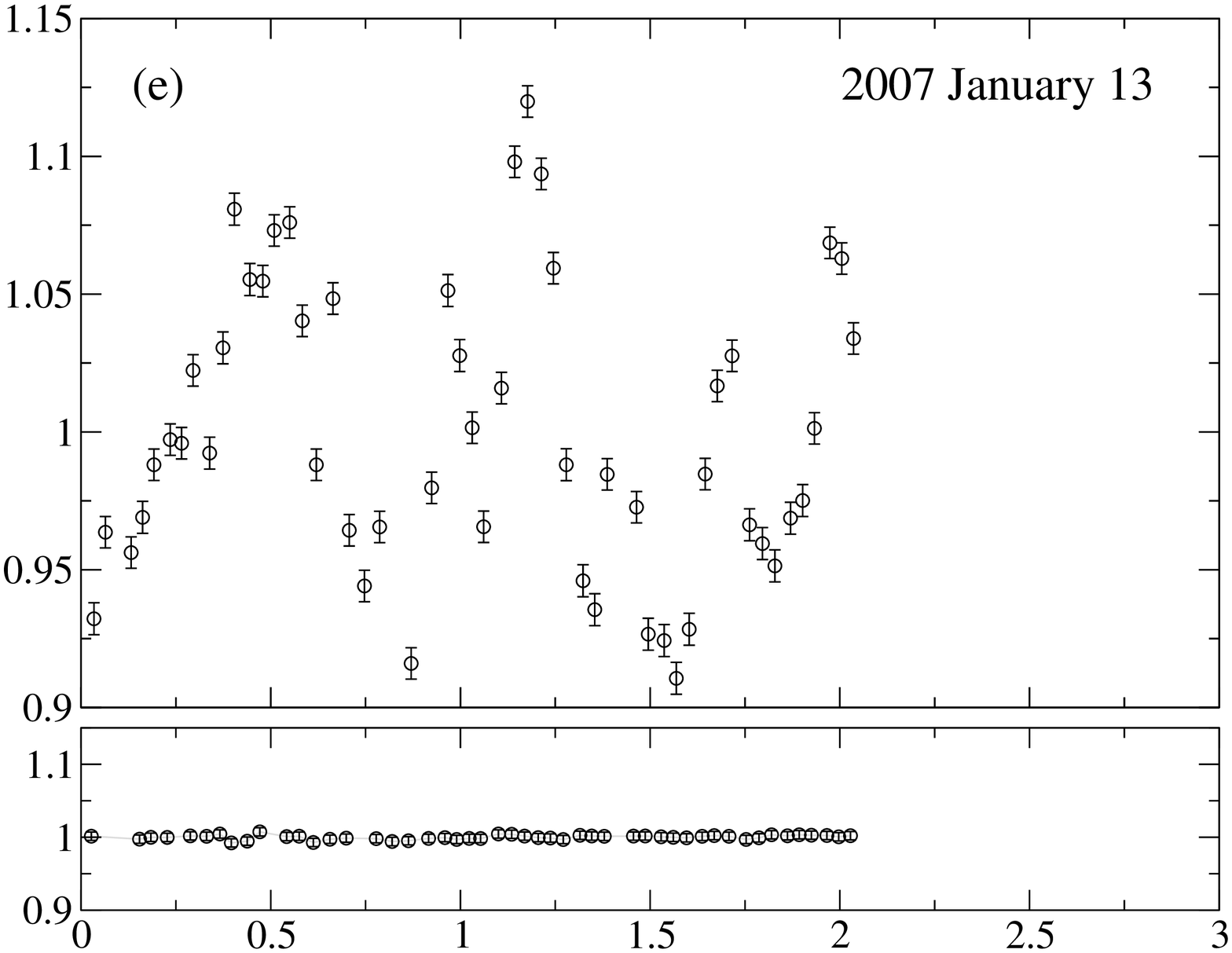}
 \end{center} 
 \end{minipage}
 \caption{4.85\,GHz light curves of J1128+5925 (upper plots) and those of the
 secondary calibrator B0836+710 (lower plots) from 2006 August (panel (a) ) until 2007
 January (panel (e) ). On the abscissa the observing time (in days) is
 displayed, the starting dates are given in the upper right corner of each
 plot. On the ordinate are the normalized flux densities in Jansky. Panels
 (a), (b) and (c) show Urumqi observations, panels (d) and (e) show Effelsberg observations.} 
 \label{fig:lcs}
\end{figure*}

In Fig. \ref{fig:lcs} the light curves of five new, previously unpublished IDV
observations of J1128+592 are
shown. As in all the previous measurements, J1128+592 showed pronounced
variability with peak-to-trough amplitudes of up to 20-25\,\%.

\begin{figure}
  \includegraphics[width=\columnwidth]{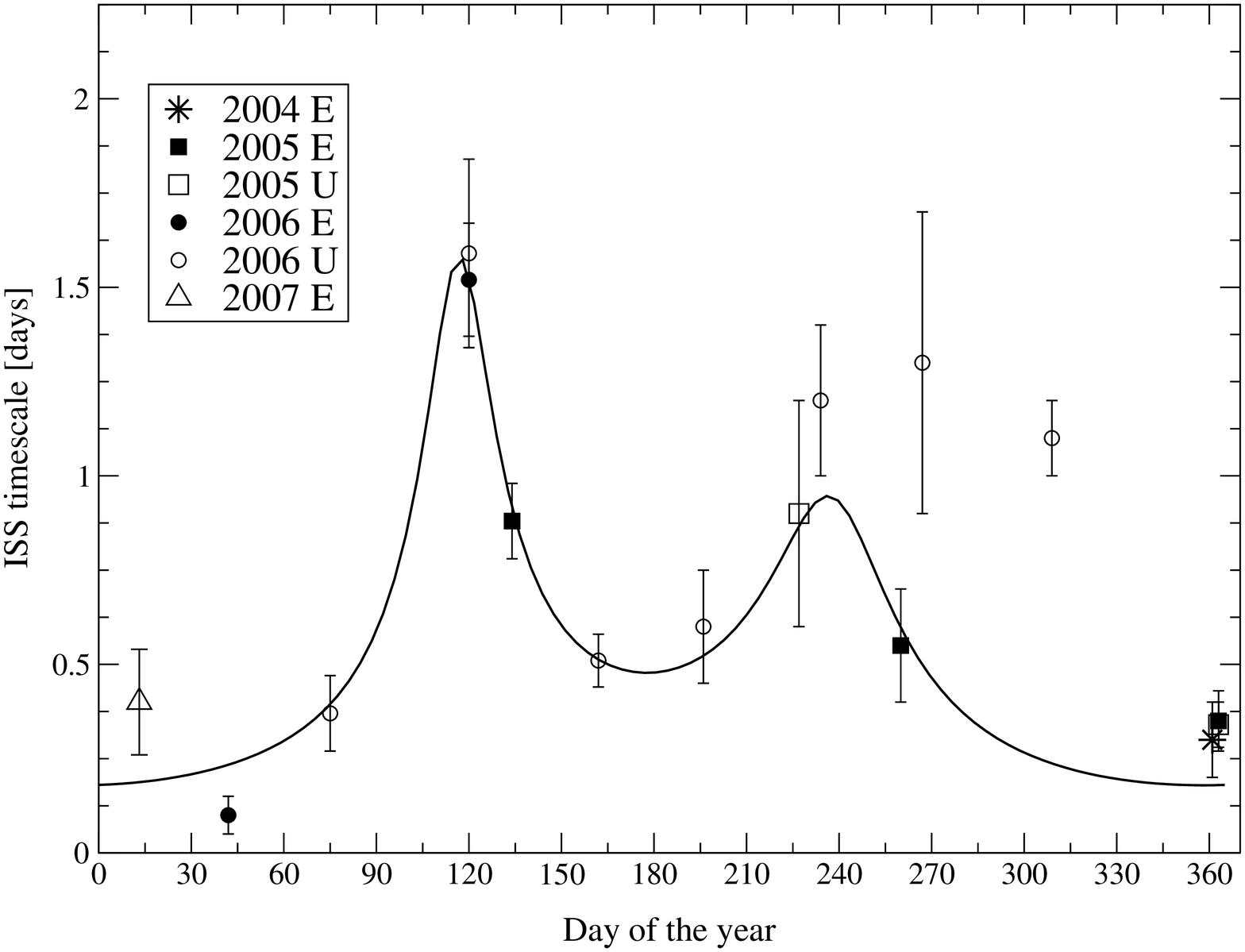}
  \caption{\label{fig:annmod} The characteristic variability time scales of
  J1128+592 plotted versus day of the year. The data are
  from 4.85\,GHz observations performed with the
  Effelsberg telescope (filled symbols) and with the Urumqi telescope (open
  symbols). Different symbols represent
  observations performed in different years: star stands for 2004, squares for
  2005, circles for 2006 and triangle for 2007. The solid curve represents the
  best fit annual modulation model as given in \cite{1128_first}, which uses 
  only the first ten observing epochs for the fit.} 
\end{figure}

The observed variability time scales of J1128+592 are significantly 
different at different dates (see Col. 5 of Table \ref{tab:obs1128}). 
In \cite{1128_first}, we proposed that the changes in the characteristic variability
time scale may be due to annual modulation. The best fit to the data was achieved
using the anisotropic annual modulation model of
\cite{bignall_newest}, which is represented by the solid curve in
Fig. \ref{fig:annmod}. 

In the anisotropic model, the scintillation time scale also depends on the
ellipticity of the scintillation pattern and on the direction in which
the relative velocity vector (between the Earth and the screen) ``cuts through'' 
the elliptical scintillation pattern.
Thus, the fitted parameters obtained from the anisotropic scintillation
model are the velocity components of the scattering screen, the scattering
length-scale (which depends on the screen distance and the scattering
angle), the angular ratio of the anisotropy and
its position angle. 

However, not all of the new observations (data points at 
day 234, day 267, day 309 and day 13) fit equally well to this model. 
It is clear that the time scales
derived from the observation in September 2005  and in September 2006 do not
agree well with each other, questioning the interpretation via annual modulation.
The larger error bar on the time scale obtained in September 2006 is
due to the fact that only one well-defined variability peak occurred
during this observation. Moreover the relatively slow time scale as measured in
November 2006 (day 309) does not agree well with the model. On the other hand, 
the acceleration of the variability time scale, which is expected after autumn
(day $\geq 240$) is basically seen again, also with the new data. Despite some 
difficulties in the determination of the time scale in December 2006 
(due to irregular sampling), the variability in this months is faster than in 
autumn. Also the measurements of January 2007 (day 13) are consistent with 
such fast variability, although this recent data point shows a bit slower
variations than in February 2006 (see Fig. \ref{fig:lcs}). 

Inspecting the structure function and autocorrelation\\function of the
2006 November data reveals a possible faster time scale of $\sim
0.5$\,day. This time scale would fit to the model. Unfortunately the sampling
and the amount of measurements obtained in the later epoch, in 2006 December are not adequate to
check the existence of two time scales. At the previous epoch (2006 September),
there is no sign of variation on an additional time scale either. More
observations are necessary to confirm or reject the appearance
of a secondary time scale. 

Apart from the time scale, the modulation index shows significant variations as well. 
Notably, in 2006 September the modulation index is half of the one measured
in the previous epoch (2006 August) and the following epoch (2006 November).
In the latter case $m=4.7$\,\%, whilst in September $m=2.3$\,\%. The annual modulation 
model cannot explain changes in the
strength of variability. These different variability indices might therefore 
suggest changes in the scattering plasma or
intrinsic changes in the source. If the scattering angle or the intrinsic
source size has increased in this epoch, we would expect a prolongation of the
time scale as well. So this might explain the discrepancy with the annual
modulation model. 

Finally, we note that the mean flux density at 4.85\,GHz changed significantly 
during the $\sim 2$ years of monitoring of J1128+592. The flux density increased by
$\sim 27$\,\% until 2006 February, then monotonously decreased 
until our last observation to approximately the same flux-density level. This long term
flux density change suggests a source-in\-trin\-sic origin (e.g. an
ejection of a new jet component), which could influence the variability
behavior of the source on IDV time scales as well. Already proposed Effelsberg monitoring observations will help us to further
investigate these questions. Additionally, proposed Very Long Baseline Array
observations can reveal any intrinsic changes in the radio structure of J1128+592.

\acknowledgements
This work is based on observations with the 100-m telescope of the MPIfR at
Effelsberg and with the 25-m Urumqi telescope of the Urumqi Observatory,
National Astronomical Observatories of the Chinese Academy of
Sciences. K.\'E. G. and N. M. have been partly supported for this research
through a stipend from the International Max Planck Research School for Radio and Infrared Astronomy at the Universities of Bonn andCologne. X.~H. Sun and J.~L. Han are supported by the National Natural Science Foundation (NNSF) of China (10521001).

\end{document}